\documentclass[conference]{IEEEtran}
\usepackage{amsmath,amssymb,graphicx,url}
\usepackage{algorithm}
\usepackage{algpseudocode}

\makeatletter
\newcommand{\linebreakand}{%
  \end{@IEEEauthorhalign}%
  \hfill\mbox{}\par
  \mbox{}\hfill\begin{@IEEEauthorhalign}%
}
\makeatother

\begin{document}

\title{Barter: Trading KV Retention for Workflow Progress in Tool-Interleaved LLM Serving}

\author{
\IEEEauthorblockN{Longji Yao}
\IEEEauthorblockA{
Sun Yat-sen University\\
Guangzhou, China\\
yaolj25@mail2.sysu.edu.cn
}
\and
\IEEEauthorblockN{Xuan Mo}
\IEEEauthorblockA{
Sun Yat-sen University\\
Guangzhou, China\\
moxuan@mail2.sysu.edu.cn
}
\linebreakand
\IEEEauthorblockN{Jialun Li}
\IEEEauthorblockA{
Guangdong Polytechnic Normal University\\
Guangzhou, China\\
jialun.li@gpnu.edu.cn
}
\and
\IEEEauthorblockN{Weigang Wu}
\IEEEauthorblockA{
Sun Yat-sen University\\
Guangzhou, China\\
wuweig@mail.sysu.edu.cn
}
}

\maketitle

\begin{abstract}
In tool-interleaved large language model (LLM) serving, workflows alternate inference with external tool execution. Delays across dependent calls accumulate in end-to-end job completion time (JCT). Retaining key--value (KV) state during tool waits can avoid recomputing earlier context, but occupies GPU memory that other workflows need to proceed. Under contention, serving one request may therefore require reclaiming another workflow's retained state. The challenge is to balance current progress against future recomputation while tool-return times and subsequent calls remain unknown.

We present \textsc{Barter}, a serving system that jointly decides which request to serve and which retained KV to reclaim. Using known context and observed workflow and tool histories, Barter compares the progress each choice enables with the cost of displacing state needed by other workflows. It can select a different request or continue current execution when reclamation is too costly. In trace-driven evaluations of sequential coding-agent workflows at five arrival rates, Barter reduces 90th-percentile (P90) JCT by 3.8--19.5\% relative to Continuum in all tested conditions with medium, long, or variable tool waits. Mean-JCT effects depend on workload and arrival rate; both metrics improve at every rate in the variable-wait condition.

\end{abstract}

\section{Introduction}
\label{sec:introduction}
Conventional large language model (LLM) serving systems primarily schedule individual inference requests, balancing latency and throughput across concurrent execution~\cite{orca,sarathi}. Key--value (KV) cache management further improves memory utilization and reuse across requests~\cite{vllm,sglang}. These mechanisms optimize request execution, but a request is not always the unit at which an application makes progress.

In tool-interleaved LLM serving, agentic applications such as information seeking~\cite{react} and software engineering~\cite{sweagent,openhands} alternate LLM calls with external tools. A subsequent call depends on the preceding output and tool feedback, so delays across calls accumulate in end-to-end job completion time (JCT). Program-aware scheduling has begun to address this broader objective by using execution history to prioritize constituent calls~\cite{agentix}. We focus on its memory implications. Sequential coding-agent workflows provide a representative setting: successive rounds of inspection, editing and testing reuse growing context~\cite{cachewise}, yet tool results determine what the agent does next and when it finishes. The serving system therefore observes the current request without knowing the remaining workflow.

Tool interactions also separate the usefulness of KV state from its immediate use. A workflow waiting for external feedback may retain attention state that will be useful when it resumes, even though it cannot execute another LLM request yet. At the same time, a different ready request may be unable to run because that state occupies GPU memory. Reclaiming KV can make room for current execution, but the suspended workflow may have to recompute the discarded context when its next request arrives~\cite{infercept}. Thus, preserving previously computed work and making room for new work compete for the same capacity. Neither maximizing cache reuse nor admitting the next ready request necessarily yields the best workflow completion time.

Recent systems recognize parts of this interaction. Continuum combines tool-aware KV retention with workflow scheduling to preserve continuity across tool waits~\cite{continuum}; CacheWise incorporates additional KV demand and predicted reuse into scheduling and eviction~\cite{cachewise}. Yet choosing which request should run and which retained state should make room for it are closely related decisions. A request that looks favorable in isolation may require reclaiming valuable context from several other workflows, whereas another request may advance after displacing less costly state. A fixed service order followed by a separate reclamation choice need not capture this difference. Moreover, the future cost of either choice depends on uncertain tool returns and subsequent calls, rather than memory occupancy alone.

Our key insight is that \emph{memory contention creates a value exchange between current workflow progress and future continuation}. Admission should therefore be evaluated together with the KV reclamation it requires. The relevant comparison is not merely how much memory becomes available, but what progress that capacity enables and what additional computation the displaced workflows may later need. This perspective lets the cost of retained state influence service selection, rather than treating cache reclamation as a consequence of a predetermined order.

We design \textsc{Barter}, a serving system that makes these choices jointly in tool-interleaved LLM serving. It uses observed workflow history and known input work to guide which ready requests are considered, then compares the progress enabled by each candidate with the consequences of reclaiming the KV it needs. Tool-wait observations inform how retained context may be reused, while current memory demand informs the tradeoff with serving other workflows. As a result, Barter may select a different request or continue existing execution instead of reclaiming expensive state. Candidate exploration is bounded, and selected actions are checked against available physical capacity before reclamation is committed. Section~\ref{sec:background} develops the tradeoff; Section~\ref{sec:system-design} presents the decision model and system design.

\begin{samepage}
We make the following contributions:
\begin{itemize}
    \item We identify how workflow progress and KV retention are coupled in tool-interleaved LLM serving, and formulate admission and reclamation as a joint value-exchange decision.

    \item We design Barter, which evaluates the progress enabled by an admission together with its effects on other workflows, using bounded search to construct executable admission--reclamation plans.

    \item We evaluate four tool-wait distributions and five arrival rates. Relative to Continuum, Barter reduces 90th-percentile (P90) JCT by 3.8--19.5\% in all tested conditions with medium, long, or variable tool waits. Mean-JCT effects depend on workload and arrival rate, with both metrics improving at every rate under variable waits.
\end{itemize}
\end{samepage}
\section{Background and Motivation}
\label{sec:background}

\subsection{LLM Serving}
\label{sec:llm-serving}

An LLM request begins with \emph{prefill}, which processes its input token sequence (the \emph{prompt}) in parallel, followed by \emph{decode}, which generates output tokens autoregressively. Their different resource demands complicate concurrent execution: batching improves utilization, but long prefills can delay ongoing decode~\cite{splitwise,distserve}. Iteration-level batching schedules work across generation steps, while chunked prefill divides long prompts into smaller schedulable chunks~\cite{orca,sarathi}. Fine-grained overlap of GPU operations provides another way to improve serving throughput~\cite{nanoflow}.

KV caching preserves computed attention state for reuse but occupies GPU memory needed by other requests. Paged allocation organizes KV into blocks to reduce fragmentation~\cite{vllm}, while prefix caching and prefix-aware scheduling extend reuse across requests~\cite{sglang,preble}. Distributed KV stores further enable reuse beyond an engine's local GPU memory~\cite{mooncake}. A compatible cached prefix lets a request skip prefill for those tokens and process only the remainder of its prompt. These mechanisms improve request execution and reuse; serving interdependent requests within an unfinished task also requires accounting for workflow completion.

\subsection{Tool-Interleaved LLM Serving}
\label{sec:tool-interleaved-serving}

Many agentic applications alternate LLM requests with external actions, incorporating the resulting observations into subsequent prompts~\cite{react,speculativeactions}. We refer to serving requests within this feedback loop as \emph{tool-interleaved LLM serving}. These applications span software engineering, interactive tasks and multi-agent pipelines, with dependencies and reusable context across calls~\cite{sweagent,helium,kvflow}. Sequential coding-agent workflows provide a representative case: rounds of inspection, editing and testing depend on tool feedback and reuse accumulated context~\cite{sweagent,openhands,cachewise}. Within each workflow, a \emph{turn} is a complete LLM request through prefill and decode; tool feedback determines subsequent prompts and termination, so neither is known in advance.

\begin{figure}[!t]
    \centering
    \includegraphics[width=\columnwidth]{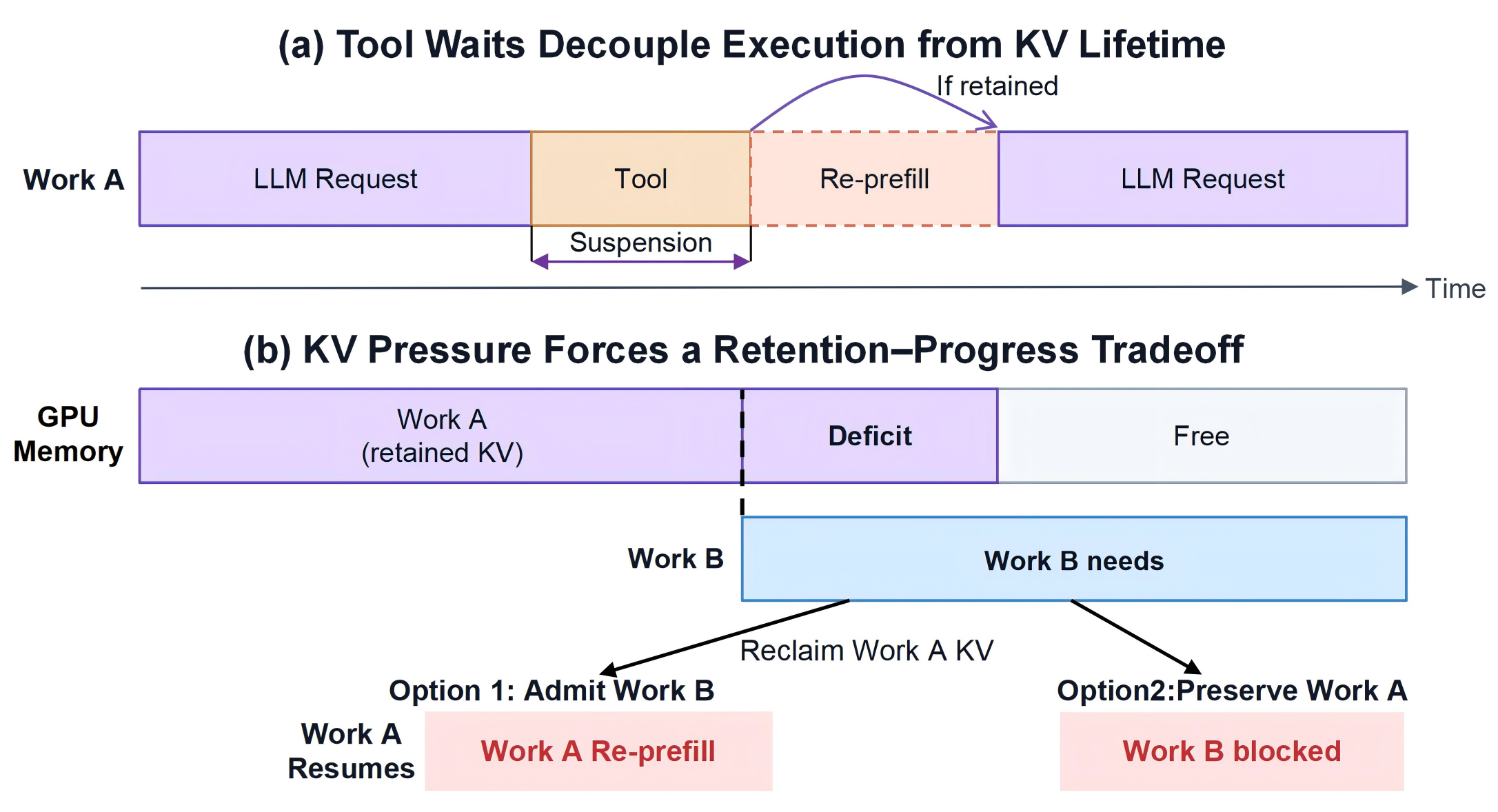}
    \caption{Tool waits and KV contention. (a) Retaining compatible KV during tool suspension avoids re-prefill on return. (b) Reclaiming A's KV admits B but may require A to reconstruct that prefix later; preserving A's KV leaves B waiting for capacity. Timelines are schematic.}
    \label{fig:tool-wait-tradeoff}
\end{figure}

During a tool call, a workflow is \emph{tool-pending}: a subsequent LLM request must await the result, although compatible KV from the preceding call may remain in GPU memory~\cite{infercept}. When the required feedback makes a subsequent request available, it becomes \emph{ready} for scheduling, whether or not that KV is still resident. Thus, execution readiness and \emph{KV residency} describe different properties. As Fig.~\ref{fig:tool-wait-tradeoff}(a) illustrates, retaining KV avoids \emph{re-prefill}---recomputing a previously processed prefix---when it is needed on return. Under memory pressure, reclaiming that KV makes blocks available to another request, but can force re-prefill when the original workflow resumes (Fig.~\ref{fig:tool-wait-tradeoff}(b)).

Delays across dependent requests accumulate in end-to-end workflow JCT. Tool execution contributes to JCT but differs from \emph{eligible waiting}, when an arrived, ready request awaits GPU service. Overlapping or speculating tool actions can reduce the former delay~\cite{conveyor,actwhilethinking}. Agentix uses program-level execution history to prioritize constituent requests~\cite{agentix}. KV management must likewise consider workflow completion, since retained state affects both current capacity and subsequent re-prefill.

Recent systems connect scheduling with KV retention. Continuum balances avoided re-prefill and renewed queueing against the \emph{holding cost} of occupying GPU memory during tool waits, then prioritizes requests by retention status and program arrival~\cite{continuum}. Brief waits can make such retention effective without prolonged occupancy. CacheWise prioritizes requests requiring fewer additional KV blocks and orders evictions by predicted reuse time~\cite{cachewise}. These approaches raise a further question: how should the cost of the \emph{particular} state displaced by an admission influence which workflow receives service?

\subsection{Motivation}
\label{sec:motivation}

We model this setting as \emph{memory-constrained scheduling of stateful workflows}. A workflow defines the completion objective, while \emph{admission} selects its current ready LLM request for GPU execution. The request's \emph{preparation cost} is the estimated time to prefill the known prompt tokens not covered by a compatible reusable KV prefix, including any re-prefill of previously computed context; it excludes subsequent output generation. Execution readiness therefore identifies candidates for service, while KV residency determines their preparation costs and additional memory requirements.

Keeping KV resident can avoid that preparation work if the context is reused, but its holding cost reflects the opportunity to use the occupied capacity for other requests. Conversely, reclamation may enable current progress at the expense of later re-prefill. Re-prefill merely restores earlier computation; useful progress requires new work beyond the recovered context. This is a \emph{value exchange between current workflow progress and future continuation}, not simply a choice to maximize cache hits or minimize eviction.

Because prefix reuse requires continuity, reclamation removes legal KV suffixes rather than arbitrary blocks. For a candidate \emph{receiver} (the request being considered for admission), its \emph{capacity deficit} is the additional physical KV blocks still required after accounting for reusable KV and \emph{ordinary capacity}, available without displacing retained state. A \emph{reclamation plan} specifies the suffixes to truncate from one or more \emph{donors} (workflows contributing retained KV). A plan is \emph{complete} only if those cuts supply exactly the deficit in distinct physical blocks while preserving contiguous reusable prefixes and respecting shared state, in-flight accesses and existing execution commitments. Different receivers can therefore incur different reclamation costs, even in the same scheduling snapshot.

\begin{figure}[!t]
    \centering
    \includegraphics[width=\columnwidth]{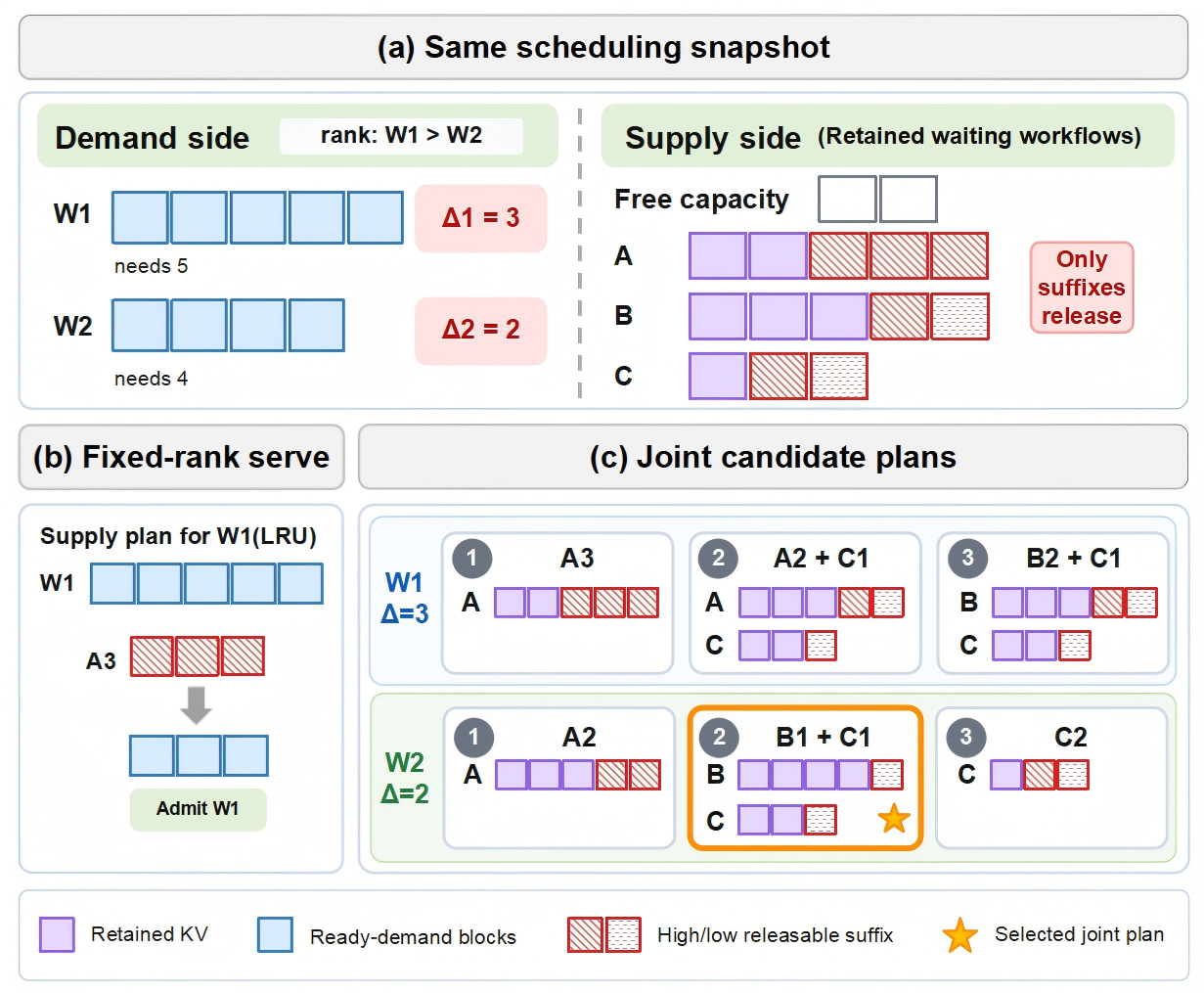}
    \caption{Why service selection depends on reclamation. (a) With two free blocks, $W_1$ and $W_2$ lack three and two, respectively. (b) Fixed rank with least-recently-used (LRU) reclamation selects $A3$ for $W_1$. (c) All shown $W_1$ plans incur high-cost reclamation, while $B1+C1$ admits $W_2$ at lower cost (star). Costs are illustrative.}
    \label{fig:joint-decision-motivation}
\end{figure}

Figure~\ref{fig:joint-decision-motivation} shows why a fixed service order can overlook the cost of its required supply. An initial \emph{reference order} prefers $W_1$ to $W_2$; LRU then reclaims $A3$ (the three-block tail of A) for $W_1$. Even with a different plan, such as $B2+C1$, admitting $W_1$ requires at least one high-cost block. In contrast, $W_2$ can fill its smaller deficit with $B1+C1$, both low-cost suffixes. Once the progress enabled and the consequences for other unfinished workflows are considered together, the cheaper plan may justify selecting $W_2$ instead.

Accordingly, a joint \emph{admission--reclamation action} is $a=(i,E)$, pairing a ready receiver $i$ with a complete legal plan $E$ of suffix truncations. The plan specifies which retained state would be displaced, rather than independently choosing arbitrary KV blocks; when no reclamation is required, $E=\emptyset$. If existing requests are already executing, continuing their service without another admission is also an available choice. For $N$ workflows, the objective is to minimize mean end-to-end JCT:
\begin{equation}
    \min_{\pi}\;\frac{1}{N}\sum_{w=1}^{N}\operatorname{JCT}_{w}(\pi),
    \qquad
    \operatorname{JCT}_{w}=t_{w}^{\mathrm{done}}-t_{w}^{\mathrm{begin}},
    \label{eq:workflow-jct-objective}
\end{equation}
where $\pi$ is an online serving policy based only on information available at decision time. The timestamps $t_w^{\mathrm{begin}}$ and $t_w^{\mathrm{done}}$ mark workflow initiation and final completion, including required tool feedback.

\noindent\textbf{Design challenges.}
Two challenges follow. First, the scheduler must estimate how an admission changes preparation work and retained memory across affected workflows, despite unknown tool outcomes and remaining workflow paths. Second, candidate receivers and their legal suffix combinations form a large search space, yet decisions must remain affordable on the scheduling path. Execution must also provide actual physical capacity and protect the selected request's complete known preparation and subsequent progress. Barter combines a common value model, bounded plan construction and capacity commitments to make these decisions, as developed next.

\section{Barter System Design}
\label{sec:system-design}

Barter realizes the value exchange identified in \S\ref{sec:motivation} by evaluating the progress enabled by an admission together with the retained KV it must displace. A reference order guides candidate exploration; a common value model compares complete admission--reclamation actions. This separation supports bounded candidate exploration without treating request selection and KV reclamation as independent decisions.

\subsection{System Overview}
\label{sec:system-overview}

\begin{figure*}[t]
    \centering
    \includegraphics[width=\textwidth]{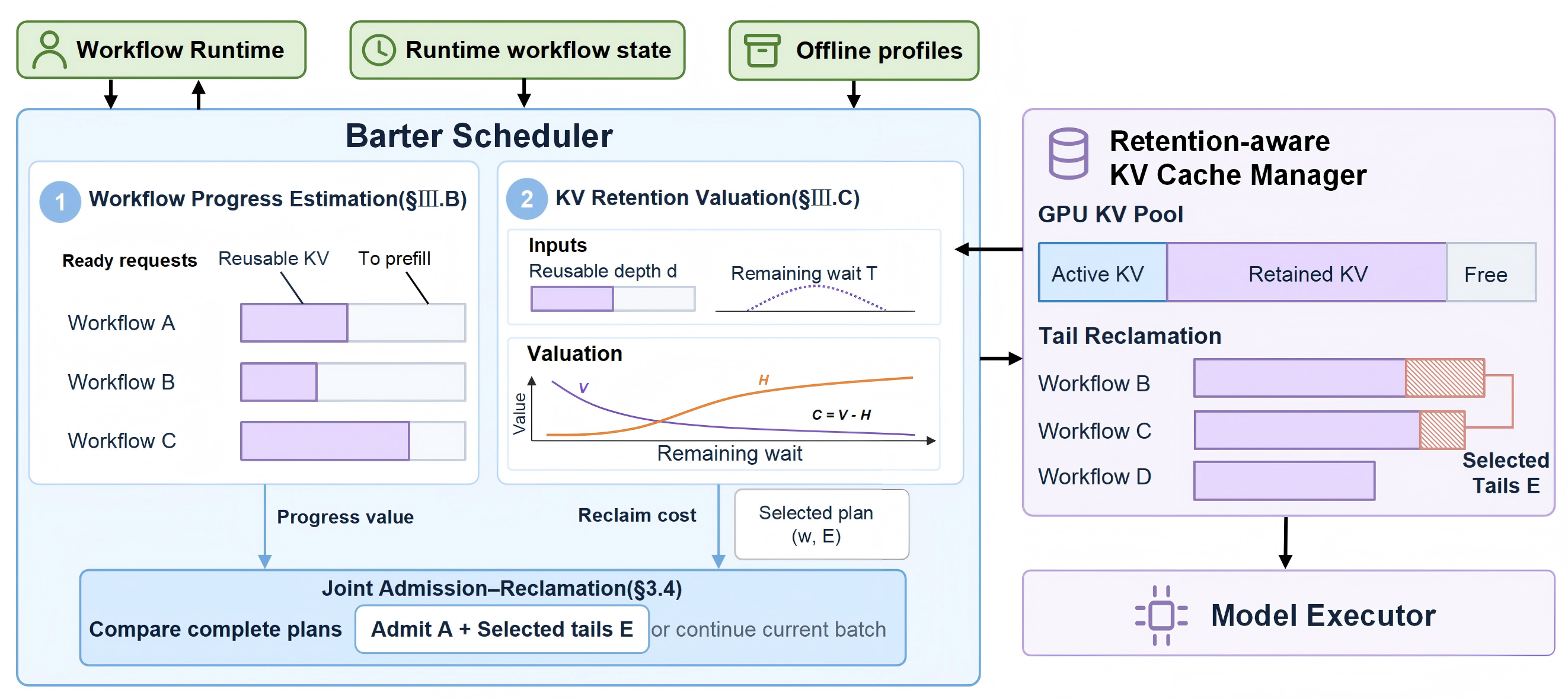}
    \caption{Barter architecture. An external workflow runtime supplies requests and lifecycle events; inside the serving engine, progress estimation and retention valuation inform joint admission and reclamation. Barter's retention-aware KV cache manager verifies and commits the chosen plan through native KV management. The valuation inset illustrates the retention tradeoff; complete actions use the common value model defined in \S\ref{sec:retention-valuation}.}
    \label{fig:barter-overview}
\end{figure*}

Figure~\ref{fig:barter-overview} distinguishes an external workflow runtime from Barter's serving-side components. The runtime launches workflows and issues successive LLM requests after their tool dependencies are satisfied. It supplies workflow identifiers and lifecycle events to the serving engine, where the Barter Scheduler tracks readiness and decides admission. To support these decisions, Barter extends the engine's existing, or \emph{native}, KV management with a \emph{retention-aware KV cache manager}. It exposes reusable prefixes, available capacity, and legal suffix-reclamation options to the scheduler; physical allocation and release remain under native control.

The scheduler contains three cooperating functions. \emph{Workflow progress estimation} maintains a reference order from known preparation work and observed service history (\S\ref{sec:progress-estimation}). \emph{KV retention valuation} assesses the effects of displaced state using the preparation profile and conditional tool-wait information (\S\ref{sec:retention-valuation}). \emph{Joint admission and reclamation} evaluates the resulting complete actions against continued execution (\S\ref{sec:joint-scheduling}). Once a plan is selected, the cache manager verifies its physical feasibility and commits the required capacity before model execution. Subsequent execution and lifecycle events update the next decision.

\subsection{Workflow Progress Estimation}
\label{sec:progress-estimation}

A workflow progresses through complete LLM turns, each comprising input preparation and output generation. Completing a turn enables the next tool interaction or workflow termination. Because future calls and the remaining output are unknown, Barter draws on three observable quantities: current preparation cost, accumulated eligible waiting, and useful service from completed turns.

For a ready request $i$, let $\ell_i$ be its known prompt length and $d_i$ its valid reusable-prefix depth. The offline execution profile determines
\begin{equation}
    K_i=T_{\mathrm{prepare}}(d_i,\ell_i),
    \label{eq:known-preparation-cost}
\end{equation}
which covers re-prefill of missing KV and preparation of new prompt tokens under the engine's chunking limits. Native capacity accounting separately determines how many blocks are required. A reusable prefix lowers $K_i$ rather than receiving an additional priority bonus.

Barter preserves scheduling delay across turns. Let $e_w(u)$ equal one when workflow $w$ has an arrived request eligible for service but waiting at time $u$, and zero otherwise. Its cumulative eligible waiting is
\begin{equation}
    W_w(t)=\int_{t_w^{\mathrm{begin}}}^{t}e_w(u)\,du.
    \label{eq:workflow-eligible-waiting}
\end{equation}
This measure includes waiting for re-prefill but excludes tool waits and time in service. Neither request completion nor tool return resets it. To account for work already received, Barter also maintains
\begin{equation}
    S_w(t)=\sum_{r\in\mathcal C_w(t)}s^{\mathrm{use}}_{w,r},
    \label{eq:completed-useful-service}
\end{equation}
where $\mathcal C_w(t)$ contains completed turns and $s^{\mathrm{use}}_{w,r}$ is their useful service, estimated from actual new input work and observed output through the frozen profile. Re-prefill adds no useful-service credit. An unfinished turn likewise does not increase $S_w$, although its actual progress reduces the current preparation requirement.

With a positive reference scale $\tau$, Barter defines
\begin{equation}
    \omega_w=1+\frac{W_w}{S_w+\tau},\qquad X_i=K_i+\tau,
    \label{eq:workflow-service-weight}
\end{equation}
and explores ready requests in decreasing order of
\begin{equation}
    P_i=\frac{\omega_{w(i)}}{X_i}
       =\frac{1+W_{w(i)}/(S_{w(i)}+\tau)}{K_i+\tau},
    \label{eq:reference-priority}
\end{equation}
where $w(i)$ identifies request $i$'s workflow. Smaller preparation costs improve the reference priority, whereas waiting receives greater weight when less useful service has been completed. This \emph{reference priority} determines exploration order, not the value of an exchange. The common value model uses the same workflow weights and preparation costs to assess both the selected request and the state it displaces. A higher-ranked request need not therefore be admitted first.

\subsection{KV Retention Valuation}
\label{sec:retention-valuation}

Reclaiming retained KV may increase the preparation required when its owner resumes. If a legal plan $E$ reduces workflow $j$'s reusable-prefix depth from $d_j$ to $d_j^E$, the additional known preparation is
\begin{equation}
    \Delta K_j(E)
       =T_{\mathrm{prepare}}(d_j^E,\ell_j)
        -T_{\mathrm{prepare}}(d_j,\ell_j).
    \label{eq:reclamation-preparation-increment}
\end{equation}
The prompt endpoint and preparation profile remain fixed. For a tool-pending workflow, the endpoint covers context known at suspension, not unseen tool feedback.

\noindent\textbf{Conditional returns and preparation value.}
The cost of losing a prefix depends on its possible reuse and on other workflows awaiting service. Let $T_j$ be the remaining time until a pending interaction ends, by return or termination, and $Y_j$ indicate a return with another LLM request. Barter conditions the empirical $(T_j,Y_j)$ distribution on the observed tool and elapsed tool-wait time, obtaining
\begin{equation}
    q_j=\Pr(Y_j=1\mid\mathcal I_s),\qquad
    \kappa_j=\mathbb E[e^{-T_j/h}Y_j\mid\mathcal I_s],
    \label{eq:conditional-return-terms}
\end{equation}
where $\mathcal I_s$ is the information available in decision snapshot $s$ and $h>0$ is a fixed time scale. For ready or committed requests, $q_j=\kappa_j=1$. Return probability $q_j$ weights an owner's remaining preparation, while $\kappa_j$ provides a time-sensitive weight for interactions with other requests.

Let $\omega_j$ abbreviate the weight of the workflow represented by entry $j$, and retain $X_j=K_j+\tau$ from \S\ref{sec:progress-estimation}. Barter evaluates the set $\mathcal C_s$ of known preparation states through
\begin{equation}
\begin{split}
    \Phi_s(\mathbf X)={}&\sum_{j\in\mathcal C_s}q_j\omega_j X_j\\
     &+\sum_{j<k}\kappa_j\kappa_k
       \min\{\omega_kX_j,\omega_jX_k\}.
\end{split}
\label{eq:shared-preparation-potential}
\end{equation}
The set represents ready requests, retained context from tool-pending workflows, and known residual preparation for requests committed to execution. A pending entry contains only the context known at suspension, not a predicted next request. The first term estimates weighted preparation work; the pairwise term approximates competition among workflows. Changes to a receiver, its donors, and other requests that reuse newly computed KV enter one projected state, so their interactions are evaluated together.

\noindent\textbf{Holding cost.}
Retained KV also occupies capacity while its workflow is tool-pending. For a projected state $\hat s$, Barter uses a pressure-sensitive holding-cost estimate,
\begin{equation}
\begin{split}
    \mathcal H_s(\hat s)={}&\lambda_s
      \sum_{j\in\mathcal P_s} b_j^{\mathrm{priv}}(\hat s)\,\mu_j,\\
    \mu_j={}&\mathbb E\!\left[\int_0^{T_j}e^{-u/h}\,du
                   \ \middle|\ \mathcal I_s\right],
\end{split}
\label{eq:holding-proxy}
\end{equation}
where $\mathcal P_s$ contains tool-pending entries, $b_j^{\mathrm{priv}}$ counts distinct reclaimable private blocks, and $\lambda_s$ reflects current demand for KV capacity. The conditional moments and pressure coefficient remain fixed within a comparison. This term estimates the cost of keeping private KV idle during the current tool wait. A tool return updates the next decision snapshot; actual capacity availability is checked by native KV management.

Combining preparation and holding effects gives the common state-cost estimate
\begin{equation}
    \Psi_s(\hat s)=\Phi_s(\mathbf X(\hat s))+\mathcal H_s(\hat s).
    \label{eq:joint-state-cost}
\end{equation}
A lower $\Psi_s$ represents less modeled preparation and holding cost. A reclamation plan changes it through the complete resulting state. The cache manager provides legal whole-block suffix cuts that preserve prefix continuity and protect shared state, in-flight accesses, and existing commitments. The next step compares these effects with the progress enabled by admission.

\subsection{Joint Admission and Reclamation}
\label{sec:joint-scheduling}

\noindent\textbf{Capacity and complete plans.}
At a decision point, Barter first obtains the \emph{mandatory execution} already required by the native engine, denoted $m$. A candidate joint action adds a ready request $i$ to this work. Before any KV modification, the engine checks the \emph{whole} action, including existing capacity commitments, blocks reused by its requests, and $i$'s complete known preparation. Let $n_i$ be the whole action's remaining physical-block requirement and $o_i$ its available ordinary capacity. The residual \emph{capacity deficit} is
\begin{equation}
    \delta_i=\max\{0,n_i-o_i\}.
    \label{eq:admission-deficit}
\end{equation}
Writing the mandatory action $m$ implicitly, a complete admission--reclamation action $a=(i,E)$ pairs the joint execution with a legal set of suffix cuts satisfying
\begin{equation}
    E\in\mathcal E_i(s),\qquad |\mathcal R(E)|=\delta_i,
    \label{eq:complete-reclamation-plan}
\end{equation}
where $\mathcal E_i(s)$ is the native-feasible plan domain and $\mathcal R(E)$ contains distinct releasable physical blocks. The receiver's own reusable KV reduces demand but is never counted again as supply. When $\delta_i=0$, $E=\emptyset$; otherwise every cut respects prefix continuity, current readers and existing commitments. An executable mandatory-only action is independently checked against the same pre-state and provides the baseline when optional admission does not proceed.

\noindent\textbf{Progress and complete post-states.}
For timing, Barter uses the next execution quantum---the chunk or decoding step permitted by the engine. For valuation, it projects completion of the receiver's full known preparation covered by the capacity commitment. Mandatory requests retain their known residual preparation after the proposed native work. The projected state includes effects on the receiver, its donors, and already-arrived requests that can reuse compatible KV produced by that preparation. If both reclamation and such reuse affect workflow $j$, its effective compatible prefix and remaining preparation are
\begin{equation}
    \widehat d_j^{\,a}=\max\{d_j^E,\widehat d_j^{\,\mathrm{new},a}\},\qquad
    \widehat K_j^{\,a}=T_{\mathrm{prepare}}(\widehat d_j^{\,a},\ell_j),
    \label{eq:joint-post-state}
\end{equation}
where both depths refer to compatible, contiguous prefixes and $\widehat d_j^{\,\mathrm{new},a}=0$ when no new reusable prefix is projected. The maximum merges two routes to the same reusable state without counting their benefits twice; actual reuse is verified by native KV management.

\begin{figure}[t]
    \centering
    \includegraphics[width=\columnwidth]{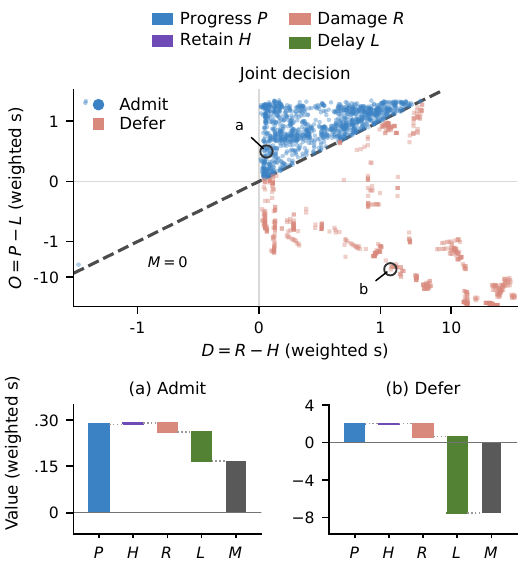}
    \caption{Value decomposition from a recorded medium tool-wait diagnostic. Top: 1,800 native-certified proposals requiring KV reclamation, separated by the $M=0$ boundary. Bottom: admitted (a) and deferred (b) examples, with $M=P+H-R-L$. Contributions are in weighted seconds and describe local decisions.}
    \label{fig:joint-value-tradeoff}
\end{figure}

\noindent\textbf{Relative action value.}
Barter evaluates complete alternatives from the same pre-state. Branch $B$ follows the mandatory native execution; branch $A$ adds $i$ with its complete reclamation plan and mixed execution. Branch $C$ keeps $A$'s execution and useful progress but preserves the displaced KV using virtual extra capacity. Thus $C$ isolates retention effects for valuation, without granting a physical action. Let $\widehat s_Z$ be branch $Z$'s projected preparation state, and $t_A,t_B$ the profiled next-batch durations. The relative cost combines execution delay with the common state-cost estimate:
\begin{equation}
\begin{split}
    J_A-J_B={}&\Omega_s e_i(t_A-t_B)\\
               &+\Psi_s(\widehat s_A)-\Psi_s(\widehat s_B),
\end{split}
\label{eq:joint-action-cost}
\end{equation}
where $\Omega_s$ sums affected eligible and committed workflow weights. With $g_i>0$ the profiled first-quantum duration, $K_i^{\mathrm{after}}$ is the preparation remaining after that quantum. The factor $e_i=1+K_i^{\mathrm{after}}/g_i$ extends its marginal execution delay to the full known preparation under a local efficiency approximation; $e_i=1$ when no preparation remains. This projection covers the known prompt, not future turns or output lengths. Costs are in weighted reference-time units. The net reclamation cost, admission opportunity, and action margin are
\begin{align}
    C_i(E)&=J_A-J_C, \label{eq:joint-reuse-damage}\\
    O_i(E)&=J_B-J_C, \label{eq:joint-opportunity}\\
    M_i(E)&=O_i(E)-C_i(E)=J_B-J_A.
    \label{eq:joint-action-difference}
\end{align}
A positive $M_i$ favors the joint action over mandatory execution under the local model. The signed $C_i$ includes additional re-prefill and the estimated holding cost avoided by reclaiming selected blocks; it need not be positive. All affected preparation states enter $\Psi_s$ together, so independently evaluated donor suffixes cannot simply be summed.

Figure~\ref{fig:joint-value-tradeoff} decomposes a recorded local margin as $M=(P-L)-(R-H)$. In this figure, $P$ denotes preparation benefit, $H$ avoided holding cost, $L$ execution delay, and $R$ re-prefill effects; $P$ is distinct from the reference priority $P_i$. The illustrated admitted and deferred decisions differ because a favorable progress opportunity can still require a costly complete plan.

\begin{algorithm}[t]
\caption{Bounded joint admission and reclamation}
\label{alg:barter-joint}
\footnotesize
\begin{algorithmic}[1]
\Require Native state $s$; shared optional-search budget $B$
\State Freeze pre-state, reference order, and value frame
\State $m\gets\Call{NativeBaseline}{s}$ \Comment{Mandatory/necessary service}
\If{$m=\bot$} \State \Call{NecessaryOrWait}{s}; \Return \EndIf
\State $f\gets\Call{OrdinaryAlternative}{s,m}$ \Comment{Full read-only plan}
\State $r_f\gets\Call{RelativeValue}{s,m,f}$ if $f\ne\bot$, else $0$
\State $\mathcal Q\gets\emptyset$
\For{$i$ in bounded reference order, excluding requests in $m,f$}
    \State $\delta_i\gets\Call{NativeWholeDeficit}{s,m\oplus i}$
    \State \textbf{continue if} $\delta_i$ is unresolved
    \For{$E\in\Call{CompletePlans}{s,m\oplus i,\delta_i,B}$}
        \State $a\gets(m\oplus i,E)$; $\hat s_a\gets\Call{ProjectPostState}{s,a}$
        \State $r_a\gets\Call{RelativeValue}{s,m,a,\hat s_a}$
        \If{$r_a<r_f$}
            \State Add $(a,r_a)$ to $\mathcal Q$; \textbf{break} \Comment{Only $i$}
        \EndIf
    \EndFor
    \State \textbf{break if} $B$ is exhausted
\EndFor
\State $\mathcal Q\gets\Call{StableCostOrder}{\mathcal Q}$; append $f$ if present
\For{$a$ in $\mathcal Q$}
    \State $v\gets\Call{FreshWholeNativeTryCommit}{s,a}$
    \If{$v=\mathrm{COMMITTED}$} \Return \EndIf
    \If{$v=\mathrm{STALE}$} \State \Call{Resnapshot}{}; \Return \EndIf
\EndFor
\State \Call{ContinueOrNecessaryNativeService}{s,m}
\end{algorithmic}
\end{algorithm}

\noindent\textbf{Bounded selection and execution.}
Reference priority determines which ready candidates are examined, not which complete action wins. For a candidate deficit, value-guided and native-order constructions produce a small set of complete plans; each is evaluated against the same mandatory baseline. A complete, acceptable plan can end \emph{that candidate's} supply search, while comparisons with other receivers continue within a shared work budget. An ordinary alternative $f$ is a complete, read-only plan certified by the native engine that admits a request without displacing retained state. When available, it sets the \emph{admission gate}: a displacing admission must have a lower relative cost, with ties favoring ordinary service. Otherwise the mandatory-only action is the baseline. When no GPU work is executing or in flight, \emph{necessary service} starts a feasible ready request through the native engine and supplies the baseline. Each candidate uses the same mandatory baseline and the local preparation scaling defined above. Search exhaustion leaves unexamined plans unresolved and does not interrupt native progress.

Algorithm~\ref{alg:barter-joint} follows the four stages above. The native baseline $m$ denotes actual mandatory execution or an executable necessary-service action, while $f$ is an optional certified ordinary alternative. \textsc{CompletePlans} returns fully constructed exact-deficit plans; an unresolved deficit or budget exhaustion never certifies infeasibility. \textsc{RelativeValue} returns the relative cost $r_a=J_A-J_B=-M_i(E)$, so lower values are better and $m$ has cost zero. The threshold $r_f$ preserves ordinary service when it is available. \textsc{FreshWholeNativeTryCommit} first certifies the entire action, then rechecks each grant against preceding committed changes. Staleness detected before the first grant triggers a new snapshot. When no optional action is selected, native continuation or necessary service proceeds. \S\ref{sec:implementation} details these execution boundaries.


\section{Implementation}
\label{sec:implementation}

We implement Barter in vLLM's scheduler and native KV management. The integration connects workflow state to the joint decision in \S\ref{sec:system-design}, while allocation and model execution remain under the engine's control.

\noindent\textbf{Workflow integration.}
Successive requests carry a common workflow identity. Arrival, request completion, tool suspension and return update the workflow's readiness and retained context. Accumulated eligible waiting persists across turns, while useful service is credited only for completed turns. During a tool wait, eligible waiting stops accumulating, but the workflow's retained context remains associated with its later continuation. Request completion and workflow completion remain separate events, allowing the runtime to account for the last external interaction before final retirement. These events also invalidate affected decision metadata, so candidate evaluation reflects the current ready set and retained state.

\noindent\textbf{Retention and capacity commitments.}
The retention-aware KV cache manager tracks content identities, physical blocks and their references. It exposes suffix-reclamation options that preserve a contiguous reusable prefix and exclude blocks protected by sharing, in-flight access or an existing execution commitment. Removing a logical reference supplies no capacity while another reference or reader still needs the block. Native checks count distinct reclaimable blocks and exclude the receiver's reusable KV from supply. Before reclamation, the manager verifies the selected plan against current block identities and reference state, then secures the selected request's complete known input capacity. This commitment survives partial prefill, although each execution step follows the engine's chunking limits. Only computed blocks become reusable cache entries. Execution receipts retire the commitment after full known preparation and progress beyond the previously completed work, or on request completion or cancellation.

\noindent\textbf{Profiles and tool histories.}
Preparation costs are queried from an offline profile for the target model and execution configuration. Tool-wait estimates use immutable empirical tables indexed by the command identified in a completed LLM output. Tables are loaded at engine startup; suffix statistics support conditioning on elapsed tool-wait time without traversing the full distribution at each decision. Return and termination observations retain separate counts; an interaction still awaiting feedback remains tool-pending. The scheduler uses the known context and observed history, with no access to future tool results or output lengths.

\noindent\textbf{Bounded online work.}
Candidate evaluation reuses preparation curves, donor-suffix metadata and reclamation-plan structures within a decision snapshot. Value estimates guide construction; each complete plan is then evaluated with the common state-cost model, including interactions among affected workflows. A shared per-round work budget accounts for discovery, valuation and plan construction across candidates. Budget exhaustion leaves unexplored alternatives unresolved, while mandatory execution and necessary service continue through the native engine. Physical or content changes invalidate reusable metadata, and the KV manager revalidates the selected plan before commitment. Cached estimates therefore guide selection, while current native checks authorize execution. The budget bounds optional search; each \texttt{schedule()} call also performs metadata maintenance, complete-plan validation and required native operations as needed. Section~\ref{sec:evaluation-cost} reports mean CPU time per call alongside end-to-end completion results.

\section{Evaluation}
\label{sec:evaluation}

We evaluate Barter on sequential coding-agent workflows, comparing end-to-end workflow JCT with vLLM and Continuum across tool-wait distributions and arrival rates. Mean JCT measures the objective in \S\ref{sec:motivation}; P90 JCT captures tail completion time. We also examine admission and reclamation through controlled variants and report mean CPU time per \texttt{schedule()} call.

\subsection{Experimental Setup}
\label{sec:evaluation-setup}

\noindent\textbf{Testbed.}
We run experiments on a server with two 12-core Intel Xeon E5-2680 v3 processors at 2.50\,GHz and approximately 252\,GiB of host memory. Each serving instance uses one NVIDIA RTX 3090 with 24\,GiB of GPU memory and runs Llama-3.1-8B-Instruct in BF16. Prefix caching, chunked prefill and native CUDA graph execution are enabled. We match the context limit, batching limits and GPU KV capacity across systems, keeping these resource settings fixed throughout the arrival-rate sweep.

\noindent\textbf{Baselines.}
We compare Barter with vLLM 0.16.0~\cite{vllm} and Continuum~\cite{continuum}. vLLM uses native FCFS request scheduling and prefix-cache eviction, providing a request-centric reference without tool-specific KV reservation. Continuum combines workflow-level arrival ordering with time-to-live (TTL) retention. It weighs avoided re-prefill and renewed queueing against memory occupancy, retaining KV through predicted tool suspensions and prioritizing returning requests whose KV remains pinned. Barter and both baselines use the same tool-parsing rules and identical replayed tool-wait distributions for each experimental condition.

\noindent\textbf{Workloads.}
We replay sequential coding-agent trajectories from SWE-bench tasks, which interleave LLM requests with file inspection, editing and testing. To fit their growing contexts on the testbed, we shorten recorded assistant and tool messages once, preserving request order, tool identities, dependencies and feedback outcomes. The resulting input and output token sequences are fixed across systems. Inference still executes on the serving engine; generation follows the recorded output tokens to control variation in model responses. Each subsequent request is submitted only after the preceding request completes and its required external feedback arrives, preserving the causal alternation between inference and tool waits.

\begin{figure}[!t]
    \centering
    \includegraphics[width=\columnwidth]{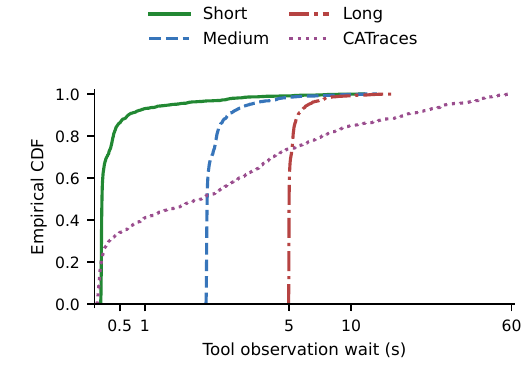}
    \caption{Empirical CDFs of tool observation waits under Short, Medium, Long and CATraces. Format-feedback events are excluded. The horizontal axis compresses the linear interval below one second and uses a logarithmic scale above it.}
    \label{fig:tool-time-cdf}
\end{figure}

We consider four tool-wait conditions. Short retains the source observation intervals. Medium and Long add fixed offsets to these intervals, placing median waits near 2 and 5 seconds while preserving their ordering. CATraces~\cite{cachewise} contributes a broader waiting-time distribution from Bash observations. We retain positive intervals shorter than 60 seconds and transfer them to the SWE trajectories through monotone quantile mapping. The mapping is fitted on the SWE training split and frozen before application to the evaluation split; Barter's training tool-time priors undergo the same transformation. Tool identities and workflow structure remain unchanged. Figure~\ref{fig:tool-time-cdf} shows the resulting distributions, separating longer typical waits from a broader tail.

For each condition, workflows arrive according to matched Poisson schedules at 0.02, 0.05, 0.075, 0.10 and 0.125 workflows/s. Increasing the rate brings the same trajectories closer together, varying their overlap and competition for KV capacity. Systems receive identical workflow arrivals, token sequences and replayed waits within each condition.

\noindent\textbf{Metrics.}
Our primary metric is end-to-end workflow JCT, defined in Eq.~\eqref{eq:workflow-jct-objective} as the interval from workflow initiation to final completion. It includes queueing, prefill, decode and external waits, including any required final tool feedback. We report arithmetic mean JCT and the 90th percentile (P90), using linear interpolation for the latter. For either statistic $S$, the relative reduction against a baseline is $1-S_{\mathrm{Barter}}/S_{\mathrm{baseline}}$. Positive reductions indicate lower JCT for Barter. For implementation cost, we report the mean CPU time per \texttt{schedule()} call separately from these end-to-end statistics.

\subsection{Overall Workflow Completion Time}
\label{sec:evaluation-overall}
\label{sec:evaluation-conditions}

\begin{figure*}[!t]
    \centering
    \includegraphics[width=\textwidth]{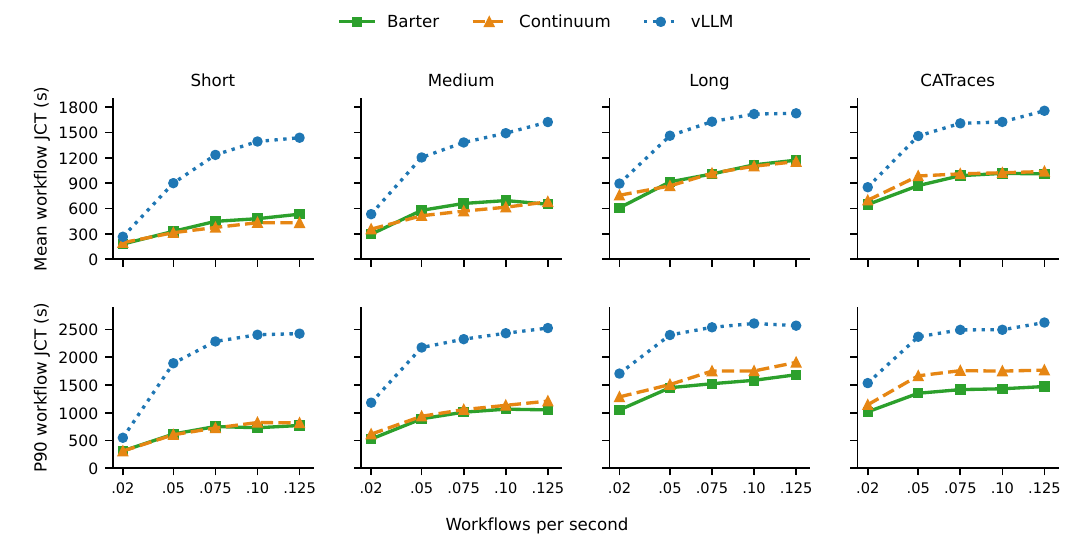}
    \caption{Mean (top) and P90 (bottom) workflow JCT for Barter, Continuum and vLLM across four tool-wait distributions and five arrival rates. Lower values are better. Barter's gains over Continuum are more consistent in P90 than in mean JCT.}
    \label{fig:reference-jct}
\end{figure*}

Figure~\ref{fig:reference-jct} compares all three systems. Barter and Continuum achieve lower mean and P90 JCT than vLLM throughout the sweep. Against Continuum, Barter lowers mean JCT in 10 of 20 distribution--rate conditions and P90 JCT in 17. All 15 Medium, Long and CATraces conditions have lower P90 JCT, with reductions of 3.8--19.5\%. P90 improvements are more consistent across conditions than mean-JCT improvements.

\noindent\textbf{Short tool waits.}
Barter lowers mean JCT by 8.2\% at 0.02 workflows/s, but raises it by 5.1--23.8\% at the other rates. P90 JCT is 1.1--3.7\% higher at the three lowest rates and 11.5\% and 5.6\% lower at 0.10 and 0.125 workflows/s. Short therefore favors Continuum in mean JCT at most rates, with a different pattern at P90. Continuum's retention and returning-request priority preserve cross-turn continuity~\cite{continuum}; brief waits limit occupancy before reuse. This may explain its stronger mean JCT at most Short rates. Section~\ref{sec:evaluation-cost} separately examines CPU time per scheduling call.

\noindent\textbf{Medium and Long tool waits.}
P90 JCT improves at every rate, by 4.4--14.5\% under Medium and 3.8--18.3\% under Long. Mean JCT does not follow the same pattern. Under Medium it falls by 16.9\% and 4.3\% at 0.02 and 0.125 workflows/s, but rises by 12.4--15.9\% at the three intervening rates. Under Long it falls by 19.7\% at 0.02 workflows/s; at the other rates, changes range from a 0.8\% reduction to a 5.5\% increase. The mean-JCT effect therefore depends on load as well as tool-wait duration. In several conditions, lower P90 JCT accompanies a higher mean, exposing different effects on the completion-time distribution.

\noindent\textbf{Variable tool waits.}
CATraces is the only distribution for which Barter lowers both statistics at every tested rate: mean JCT improves by 0.7--11.6\% and P90 JCT by 11.1--19.5\%. The largest P90 reduction is 19.5\%, or 342.14 seconds, at 0.075 workflows/s. Long and CATraces have similar mean tool waits, 5.25 and 5.86 seconds, but their medians are 5.00 and 1.89 seconds and their 99th percentiles are 9.31 and 50.51 seconds, respectively (Fig.~\ref{fig:tool-time-cdf}). This contrast shows why mean tool-wait duration alone does not characterize the serving tradeoff. CATraces provides the clearest evidence of simultaneous mean and tail benefits under heterogeneous waits.

\noindent\textbf{Load and decision tradeoffs.}
The five rates change concurrent demand while preserving workflow contents and waits. Neither mean nor P90 gains grow monotonically with arrival rate. Barter's reference priority and complete-action comparison can change which workflows advance and which incur re-prefill. One possible explanation for the mean--P90 differences is a redistribution of delay across workflows; the aggregate results do not distinguish this effect from implementation cost. The ablation below examines the admission gate and reclamation strategy separately.

\subsection{Admission and Reclamation Ablation}
\label{sec:evaluation-ablation}

We use a $2\times2$ ablation of the admission gate in \S\ref{sec:joint-scheduling} and the reclamation strategy (Table~\ref{tab:ablation-configurations}). The Value variants use Barter's reclamation strategy; the LRU variants construct reclamation plans directly in LRU order. Disabling the gate removes value-based admission rejection, including early rejection by value bounds, while retaining the common model for evaluating complete plans. All variants share the reference order, tool-wait priors, preparation profile, candidate window, search budget and native execution rules.

\begin{table}[!t]
    \centering
    \caption{Configurations for the admission and reclamation ablation.}
    \label{tab:ablation-configurations}
    \begin{tabular}{@{}lcc@{}}
        \hline
        Configuration & Admission gate & Reclamation \\
        \hline
        Full Barter & On & Value \\
        No admission gate & Off & Value \\
        LRU supply & On & LRU \\
        No gate + LRU & Off & LRU \\
        \hline
    \end{tabular}
\end{table}

The ablation compares the four variants at 0.10 workflows/s, reporting mean and P90 JCT separately. Re-prefill volume and executed-exchange records provide complementary evidence about their execution behavior. At a fixed reclamation strategy, the gate comparison tests whether rejecting costly admissions benefits workflow completion; at a fixed gate setting, Value versus LRU tests the choice of retained state to reclaim. The difference between the gate's two conditional effects measures their interaction, separately for mean and P90 JCT.

\begin{figure}[t]
    \centering
    \includegraphics[width=\columnwidth]{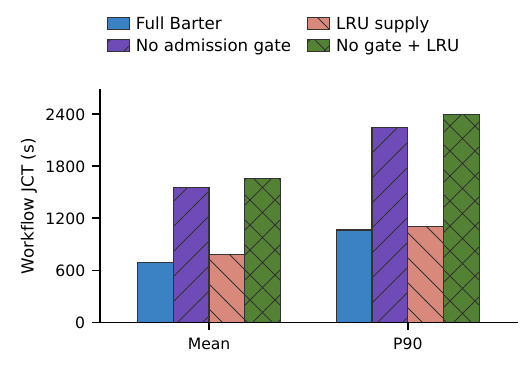}
    \caption{Impact of admission control and KV reclamation on mean and P90 workflow JCT at 0.10 workflows/s. The four configurations isolate the effects of the admission gate and reclamation strategy.}
    \label{fig:ablation}
\end{figure}




\subsection{Implementation Cost}
\label{sec:evaluation-cost}

Table~\ref{tab:implementation-cost} reports mean CPU time per \texttt{schedule()} call under Medium at 0.10 workflows/s, computed as CPU time accumulated within the function divided by the number of measured calls. Barter averages 3.055\,ms, compared with 0.483\,ms for Continuum and 0.377\,ms for vLLM---6.33$\times$ and 8.10$\times$, respectively. This metric covers scheduling and KV-management work performed within the call, including Barter's candidate discovery, valuation and plan construction.

The end-to-end JCT measurements include execution of this scheduling work. At the same condition, Barter's mean JCT is 12.8\% higher than Continuum's, while its P90 is 6.1\% lower. The higher per-call cost motivates further amortizing repeated metadata and valuation work. Its end-to-end impact also depends on call frequency and changes in batching and re-prefill, so matched JCT comparisons remain important when evaluating these optimizations.

\begin{table}[t]
    \centering
    \caption{Mean CPU time per \texttt{schedule()} call under Medium at 0.10 workflows/s.}
    \label{tab:implementation-cost}
    \begin{tabular}{@{}lr@{}}
        \hline
        System & Mean CPU time per call (ms) \\
        \hline
        vLLM & 0.377 \\
        Continuum & 0.483 \\
        Barter & 3.055 \\
        \hline
    \end{tabular}
\end{table}

\section{Related Work}
\label{sec:related-work}

\noindent\textbf{Tool-interleaved LLM serving and KV retention.}
InferCept manages context across external interactions to reduce re-prefill and GPU resource waste~\cite{infercept}. Continuum combines tool-aware retention timeouts with program-level scheduling, balancing avoided re-prefill and renewed queueing against holding costs~\cite{continuum}. CacheWise targets coding-agent sessions, prioritizing requests requiring fewer additional KV blocks and ordering evictions by predicted reuse~\cite{cachewise}. Barter evaluates each candidate admission together with its complete suffix-reclamation plan. Its common value model allows re-prefill and estimated holding costs to influence which request receives service and thus which workflow advances. At a different granularity, AgentKV uses phase-specific attention queries to select token-level KV entries for retention~\cite{agentkv}. Its attention-state compression differs from Barter's capacity allocation across workflows through suffix reclamation and subsequent re-prefill.

\noindent\textbf{Program- and graph-aware scheduling.}
Agentix prioritizes and preempts LLM calls using their programs' execution histories to reduce end-to-end program latency~\cite{agentix}. Graph-aware systems additionally exploit dependencies among agent invocations. KVFlow derives steps-to-execution from an Agent Step Graph to guide fixed-prompt prefix eviction and overlapped CPU-to-GPU prefetching~\cite{kvflow}. TOPAS jointly selects resident agent prefixes and executing requests, balancing estimated progress along each task's longest remaining service path with downstream reuse and prefix movement and preemption costs~\cite{topas}. Like Barter, TOPAS coordinates execution and residency, but its remaining-path estimates and agent-prefix sets define a different decision setting. Barter addresses tool-interleaved serving; its current implementation uses sequential workflows whose future calls and termination are unknown. It uses observed context, workflow waiting history and tool-wait distributions conditioned on elapsed tool-wait time, without estimating the remaining workflow path. Its joint action couples a ready request with the suffix-reclamation plan needed to cover that admission's capacity deficit.

\noindent\textbf{Heterogeneous execution and disaggregated serving.}
MARS coordinates GPU inference and co-located CPU tool execution through cross-resource admission control, continuation-aware scheduling and adaptive KV retention~\cite{mars}. Its control scope includes CPU tool congestion; Barter treats tool execution as an external pause and makes admission--reclamation decisions within the serving engine. Conveyor instead overlaps partial tool execution with decoding through a tool-facing interface~\cite{conveyor}. At cluster scale, Mooncake builds a distributed KV pool for disaggregated prefill and decode~\cite{mooncake}, while DualPath balances KV loading across both engine types to relieve storage-bandwidth bottlenecks~\cite{dualpath}. These systems address tool execution and cross-engine data movement. Barter focuses on whether a local admission warrants displacing retained state and which complete reclamation plan should support it, without requiring tool restructuring or disaggregated storage.


\section{Conclusion}
\label{sec:conclusion}

We presented Barter, a system for tool-interleaved LLM serving that jointly selects request admission and KV reclamation to advance workflows. A common value model compares the progress enabled with the effects on displaced workflows; bounded search constructs complete plans, and native KV management secures their execution capacity. Across the evaluated arrival rates, Barter lowers P90 JCT by 3.8--19.5\% relative to Continuum in every Medium, Long and CATraces condition. Mean-JCT effects depend on workload and arrival rate, with both metrics improving at every CATraces rate. The results show that considering workflow progress and displaced state together can improve tail completion. Reducing decision-computation costs and refining holding-cost estimates offer concrete directions for broadening the mean-JCT benefits.

\section*{Acknowledgments}
The authors used OpenAI's ChatGPT~\cite{openai_chatgpt_2026} to assist with drafting and language refinement in Sections I--VII. The authors retain responsibility for the technical content, results, and references.


\bibliographystyle{IEEEtran}
\bibliography{references}

\end{document}